\documentclass[aps,showkeys,twocolumn,preprintnumbers,amsmath,amssymb,superscriptaddress,floatfix,nofootinbib]{revtex4}
\usepackage{graphicx,color,dcolumn,booktabs,bm}
\usepackage{amsmath}
\usepackage{graphicx}
\usepackage{longtable,lscape}
\usepackage{txfonts}
\usepackage{overpic}
\usepackage{epsfig}
\usepackage{amssymb}
\usepackage{rotating}
\usepackage{epstopdf}
\usepackage{ulem}
\usepackage{appendix}
\usepackage{indentfirst}
\usepackage{feynmf}   
\usepackage{slashed}  
\usepackage{cases}
\usepackage{color}
\usepackage{multirow}
\usepackage{graphicx,color,dcolumn,booktabs,bm}
\usepackage{cases}
\usepackage{array}

\graphicspath{{Figures/}} %

\usepackage[colorlinks, citecolor=blue,anchorcolor=red,menucolor=red, linkcolor=red,filecolor=red,runcolor=red,urlcolor=blue,frenchlinks=red]{hyperref}

\begin{document}
\title{Strong decays of $P-$wave doubly charmed and bottom baryons}
\author{Ya-Li Shu}
\affiliation{Department of Physics, Hunan Normal University, Changsha 410081, China}
\affiliation{Key Laboratory of Low-Dimensional Quantum Structures and Quantum Control of Ministry of Education, Changsha 410081, China}
\affiliation{Key Laboratory for Matter Microstructure and Function of Hunan Province, Changsha 410081, China}

\author{Qing-Fu Song}
\affiliation{Department of Physics, Hunan Normal University, Changsha 410081, China}
\affiliation{Key Laboratory of Low-Dimensional Quantum Structures and Quantum Control of Ministry of Education, Changsha 410081, China}
\affiliation{Key Laboratory for Matter Microstructure and Function of Hunan Province, Changsha 410081, China}
\affiliation{School of Physics, Central South University, Changsha 410083, China}

\author{Qi-Fang L\"{u}}\email{lvqifang@hunnu.edu.cn}
\affiliation{Department of Physics, Hunan Normal University, Changsha 410081, China}
\affiliation{Key Laboratory of Low-Dimensional Quantum Structures and Quantum Control of Ministry of Education, Changsha 410081, China}
\affiliation{Key Laboratory for Matter Microstructure and Function of Hunan Province, Changsha 410081, China}

\begin{abstract}
In this work, we investigate the strong decays for $P-$wave excited states of doubly charmed and bottom baryons in the constituent quark model. Our results indicate that some $\lambda-$mode $\Xi_{cc/bb}(1P)$ and $\Omega_{cc/bb}(1P)$ states are relatively narrow, which are very likely to be discovered by future experiments. The light meson emissions for the low-lying $\rho-$mode states are highly suppressed due to the orthogonality of  wave functions between initial and final states. Moreover, the strong decay behaviors for doubly charmed and bottom baryons preserve the heavy superflavor symmetry well, where the small violation originates from the finite heavy quark masses and different phase spaces. We hope that  present theoretical results for undiscovered doubly charmed and bottom baryons can provide helpful information for future experiments and help us to better understand the heavy quark symmetry. 

\end{abstract}


\maketitle

\section{Introduction}
In the past two decades, a growing number of new hadrons in the heavy quark sector have been observed experimentally.  Besides the exotic states, some of them may belong to the conventional heavy baryons, and provide an excellent opportunity for us to investigate and establish the traditional baryon spectroscopy. In the Review of Particle Physics~\cite{ParticleDataGroup:2024cfk}, more than fifty hadrons are placed under the charmed and bottom baryons. Understanding the nature of these numerous particles and searching for the more missing heavy resonances have become an intriguing and important topic in hadron physics. 

Until now, most candidates of conventional baryons are accommodated into the singly heavy baryons, while the experimental observations for doubly heavy baryons are still scarce. In 2002, the SELEX Collaboration reported the observation of a doubly charmed baryon $\Xi_{cc}^{+}$~\cite{Mattson:2002vu,Ocherashvili:2004hi}, but the subsequent experiments and theoretical analyses did not confirm its existence~\cite{Ratti:2003ez,Aubert:2006qw,Chistov:2006zj,Aaij:2013voa}. Surprisingly, the LHCb Collaboration reported a highly significant structure $\Xi_{cc}^{++}(3621)$ in the $\Lambda_{c}^{+}K^{-}\pi^{+}\pi^{+}$ invariant mass spectrum in 2017~\cite{Aaij:2017ueg}. The mass and lifetime are then measured precisely by the LHCb Collaboration~\cite{LHCb:2018zpl,LHCb:2019qed}. Also, the LHCb Collaboration attempt to hunt for more doubly heavy baryons, such as $\Xi_{cc}^+$~\cite{LHCb:2021eaf, LHCb:2019gqy}, $\Omega_{cc}^+$~\cite{LHCb:2021rkb}, $\Xi_{bc}^+$~\cite{LHCb:2022fbu}, $\Xi_{bc}^0$~\cite{LHCb:2020iko,LHCb:2021xba} and $\Omega_{bc}^0$~\cite{LHCb:2021xba}, but no significant signal has been discovered so far. Moreover, the LHCb Collaboration has observed a doubly heavy tetraquark  $T_{cc}^{+}(3875)$~\cite{LHCb:2021vvq,LHCb:2021auc}, which may provide valuable clue for doubly heavy baryons. 

\begin{figure}[!htb]
	\centering
	\includegraphics[scale=1]{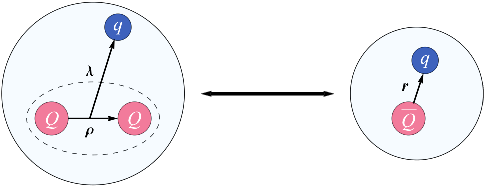}
	\caption{A sketch of heavy superflavor symmetry between doubly heavy baryons and heavy mesons. Here, the $Q$ represents the charm or bottom quark, and the $q$ stands for the light quark up, down or strange, respectively.}
	\label{sketch}
\end{figure}

Doubly heavy baryons offer an excellent platform for us to investigate the heavy quark symmetry and chiral dynamics simultaneously, where two distinct subsystems with quite different properties exist. One is the heavy-heavy subsystem, and the other is the heavy-light subsystem between a light quark and two heavy quarks. In the heavy quark limit, the two heavy quarks act as a compact color antitriplet source, and interact with the light quark degree of freedom. This is the so called heavy superflavor symmetry that  establishes a  connection between doubly heavy baryons and heavy mesons. A sketch for heavy superflavor symmetry is shown in Figure~\ref{sketch}. Also, this symmetry is broken explicitly owing to the finite heavy quark masses. Therefore, it is a important task to establish the spectroscopy of doubly heavy baryons and investigate the emergent heavy superflavor symmetry and its violation.      

Theoretically, the mass spectra, productions, and weak or radiative decays for doubly heavy baryons have been extensively studied within various approaches~\cite{Kiselev:2001fw,Ebert:1996ec,Tong:1999qs,Ebert:2002ig,Gershtein:2000nx,Roberts:2007ni,Ortiz-Pacheco:2023kjn,Valcarce:2008dr,Lu:2017meb,Yu:2022lel,Li:2022oth,Eakins:2012jk,Shah:2017liu,Soto:2020pfa,Savage:1990di,Song:2022csw,Cohen:2006jg,Wei:2015gsa,Aliev:2012nn,Liu:2009jc,Brown:2014ena,Padmanath:2015jea,Mathur:2018rwu,Mathur:2018epb,Albertus:2009ww,White:1991hz,Li:2017ndo,Yu:2017zst,Ebert:2004ck,Roberts:2008wq,Branz:2010pq,Albertus:2010hi,Qin:2021zqx,Bahtiyar:2018vub,Chen:2016spr,Cheng:2021qpd,Silvestre-Brac:1996myf,Yoshida:2015tia,Ma:2022vqf}. On the contrary, the investigations on strong decay behaviors for the low-lying doubly heavy baryons seems relatively few and unsystematic~\cite{Eakins:2012fq,Xiao:2017udy,Mehen:2017nrh,Ma:2017nik,Xiao:2017dly,Yan:2018zdt,He:2021iwx,Chen:2022fye,Song:2023cyk}. Meanwhile, the strong decay behaviors are essential to understand the internal structures for the low-lying states and investigate the similarity between doubly heavy baryons and heavy mesons. Also, the predicted decay properties of excited states are helpful for future experimental searches. Thus, it is necessary to study the strong decays of doubly heavy baryons carefully and systematically.

In previous works~\cite{Xiao:2017udy,Xiao:2017dly,He:2021iwx}, we adopted the chiral quark model or quark pair creation model to investigate the strong decays of doubly charmed and bottom baryons, but the simple harmonic oscillator wave functions were employed for calculations. Recently, we perform a systematic study for bottom-charmed baryons by combining the potential model and quark-chiral field interactions, where the masses and strong decays can be obtained within the framework of a unified quark model~\cite{Song:2023cyk,Yoshida:2015tia}. To explore the internal structures for the charmed and bottom baryons, it is time to systematically explore their strong decays by using the obtained realistic wave functions. 

In this work, we follow the similar route with bottom-charmed systems to investigate the strong decay behaviors of $P-$wave doubly charmed and bottom baryons $\Xi_{cc/bb}$ and $\Omega_{cc/bb}$. We first solve the three-body Schr\"odinger equation to get the masses and realistic wave functions, and adopt the obtained wave functions together with the quark-chiral field interactions to calculate the Okubo-Zweig-Iizuka-allowed (OZI-allowed) two-body strong decays.  Our results indicate that some $P-$wave doubly heavy charmed and bottom baryons are relatively narrow, which are very likely to be discovered by future experiments. Also, it is found that the heavy superflavor symmetry is preserved well in both charmed and bottom sectors. 

This paper is organized as follows. The formalism for pseudoscalar meson emissions is briefly introduced in Sec.~\ref{FORMALISM}. We present the numerical results and discussions of strong decays for doubly charmed and bottom baryons in Sec.~\ref{low-lying}. A summary is given in the last section.

\section{Method}{\label{FORMALISM}}

\begin{figure}[!htb]
	\centering
	\includegraphics[scale=1.0]{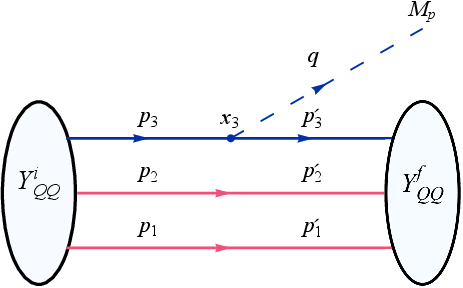}
	\caption{The pseudoscalar meson emission. $Y^i_{QQ}$ and $Y^f_{QQ}$ represent the initial and final doubly charmed or bottom baryons, respectively. $M_p$ is a pseudoscalar meson.}
	\label{emission}
\end{figure}

In the quark model, the interaction between the light quark inside a doubly charmed or bottom baryon and the pseudoscalar meson can be described by the Yukawa interaction. This interaction is considered to be the dominant contribution to one meson emission process, where a doubly charmed or bottom baryon $Y_{QQ}^{i}(P_{i})$ decays into a final  baryon $Y_{QQ}^{f}(P_{f})$ plus  a pseudoscalar meson $M_{p}(q)$ as shown in Figure~\ref{emission}. 

The coupling between the pseudoscalar meson and the light quark can be defined as
\begin{equation}
	\mathcal{L}_{M_{p}qq}=\frac{g_{A}^{q}}{2 f_{p}} \bar{q} \gamma_{\mu} \gamma_{5} \vec{\tau} q \cdot \partial^{\mu} \vec{M}_{p},   
\end{equation}
where  the $q$ stands for the quark field, $g_{A}^{q}$ is the quark-axial-vector coupling,   and $ f_{p}$ is the decay constant. In present work, $f_{\pi}$= 93 $\mathrm{MeV}$ and $f_{K}$ =111 $\mathrm{MeV}$ are adopted that have been extensively used in quark model calculations~\cite{Arifi:2022ntc,Arifi:2021orx,Nagahiro:2016nsx,Xiao:2017udy,Wang:2018fjm,Liu:2019wdr,Lu:2022puv,Song:2023cyk}. The $\vec{M}_{p}$ is a matrix of light pseudoscalar mesons and can be defined as 
\begin{equation}
	\vec{M}_{p}=\left(\begin{array}{ccc}
		\frac{1}{\sqrt{2}} \pi^{0}+\frac{1}{\sqrt{6}} \eta & \pi^{+} & K^{+} \\
		\pi^{-} & -\frac{1}{\sqrt{2}} \pi^{0}+\frac{1}{\sqrt{6}} \eta & K^{0} \\
		K^{-} & \bar{K}^{0} & -\sqrt{\frac{2}{3}} \eta
	\end{array}\right) .
\end{equation}

The wave function for the $Y_{QQ}$ ($Y_{QQ}=\Xi_{cc}$, $\Xi_{bb}$, $\Omega_{cc}$, or $\Omega_{bb}$) baryon with mass $M_{Y_{QQ}}$ in the rest frame can be expressed in the momentum representation  as
\begin{equation}
	\begin{aligned}
		\left|Y_{QQ}(J)\right\rangle=& \sqrt{2 M_{Y_{QQ}}} \sum_{\{s, l\}} \int \frac{d^{3} \boldsymbol{p}_{\rho}}{(2 \pi)^{3}} \int \frac{d^{3} \boldsymbol{p}_{\lambda}}{(2 \pi)^{3}}\\ & \frac{1}{\sqrt{2 m_{1}}} \frac{1}{\sqrt{2 m_{2}}}\frac{1}{\sqrt{2 m_{3}}} \psi_{l_{\rho}}(\boldsymbol{p}_{\rho})\psi_{l_{\lambda}}\left(\boldsymbol{p}_{\lambda}\right)\\
		&\left|q_{1}\left(p_{1}, s_{1}\right)\right\rangle\left|q_{2}\left(p_{2},s_{2}\right)\right\rangle\left|q_{3}\left(p_{3}, s_{3}\right)\right\rangle.
	\end{aligned}
\end{equation}
Then, the decay amplitude for $Y_{QQ}^{i}(P_{i})$ $\to$ $Y_{QQ}^{f}(P_{f})$+$M_{p}(q)$ can be obtained by
\begin{equation}{\label{2}}
	\begin{aligned}
		-i\mathcal{T}=&-i\frac{g_{A}^{q}g_{f}}{2 f_{p}} \sqrt{2 M_{i}} \sqrt{2 M_{f}}\int d^{3}\boldsymbol\lambda e^{i\boldsymbol{q}_{\lambda}\cdot\boldsymbol{\lambda}} \\
		& \times \Bigg\langle Y_{QQ}^{f}\Bigg|i\Bigg\{\Bigg(1-\frac{\omega}{2m_{3}}+\frac{\omega}{m_{1}+m_{2}+m_{3}}\Bigg)\boldsymbol{\sigma} \cdot \boldsymbol{q} \\ &+\frac{\omega}{m_{3}}\boldsymbol \sigma \cdot \boldsymbol{p}_{\lambda}\Bigg\}\Bigg| Y_{QQ}^{i}\Bigg\rangle,
	\end{aligned}
\end{equation}
and  the $\boldsymbol{q_{\lambda}}$ is defined as 
\begin{equation}
	\boldsymbol{q_{\lambda}}=\frac{m_{1}+m_{2}}{m_{1}+m_{2}+m_{3}}\boldsymbol{q}.
\end{equation}
The $g_{f}$ denotes the overlap of flavor wave functions,  $M_{i}$ is the mass of initial baryon, $M_{f}$ is the mass of final baryon,  $q=(\omega,\boldsymbol{q})$ is the 4-momentum of  outgoing pseudoscalar meson, $m_{i}$ is the constituent quark mass with $m_{1}=m_{2}=m_{Q}$ and $m_{3}=m_{u/d/s}$.

The masses and wave functions for initial and final baryons are obtained from the potential model~\cite{Yoshida:2015tia}. The measured mass of $\Xi_{cc}^{++}(3621)$ are used to adjust the zero point energy, and the details of calculating mass spectra  and wave functions for doubly heavy baryons can be found in the original work~\cite{Yoshida:2015tia}. 
Here, we list the calculated masses for low-lying doubly charmed and bottom baryons in Table~\ref{mass1} and~\ref{mass2}, and the predictions of other quark models are also shown for comparisons. Meanwhile, the wave functions of these states also obtained, which can be employed for calculations of strong decays.

\begin{table*}[!htbp]
	\begin{center}
		\caption{\label{mass1} The mass spectra of $\Xi_{cc}$ and $\Omega_{cc}$ in MeV.}
		\renewcommand{\arraystretch}{1.5}
		\normalsize
		\begin{tabular*}{18cm}{@{\extracolsep{\fill}}p{1.9cm}<{\centering}p{0.2cm}<{\centering}p{0.2cm}<{\centering}p{0.2cm}<{\centering}p{0.2cm}<{\centering}p{0.2cm}<{\centering}p{0.2cm}<{\centering}p{0.2cm}<{\centering}p{0.2cm}<{\centering}p{0.8cm}<{\centering}p{0.8cm}<{\centering}p{0.8cm}<{\centering}p{0.8cm}<{\centering}p{0.8cm}<{\centering}p{0.8cm}<{\centering}}
			\hline\hline
			States& $n_{\rho}$ & $n_{\lambda}$ & $l_{\rho}$ & $l_{\lambda}$ & $S_{\rho}$ & $J_{\rho}$ & $j$	&  $J^P$&Mass&\cite{Ortiz-Pacheco:2023kjn}&\cite{Roberts:2007ni}&\cite{Lu:2017meb}&\cite{Yu:2022lel}&\cite{Ebert:2002ig}\\
			\hline
			$\Xi_{cc} (1S)$	  &	0	&	0	&	0	&	0	&	1   &1    &$\frac{1}{2}$	&	$\frac{1}{2}^+$  & 3621&3619 &3674&3606&3640&3620\\
			$\Xi_{cc}^{*}(1S)$      &	0	&	0	&	0	&	0	&	1   &1    &$\frac{1}{2}$	&	$\frac{3}{2}^+$    & 3690&3686&3753 &3675&3695&3727\\ 
   $\Tilde{\Xi}_{cc} (\frac{1}{2}^-, \frac{1}{2})$&	0	&	0	&	1	&	0	&	0   &   1  &  $\frac{1}{2}$	&   $\frac{1}{2}^-$ &	3883&3823&3910&3873&3932& 3838\\
			$\Tilde{\Xi}_{cc} (\frac{3}{2}^-, \frac{1}{2})$&	0	&	0	&	1	&	0	&	0   &   1  &  $\frac{1}{2}$	&   $\frac{3}{2}^-$  &	3885&3855&3921&3916&3978 & 3959\\
			$\Xi_{cc} (\frac{1}{2}^-, \frac{1}{2})$&	0	&	0	&	0	&	1	&	1   &   1  &  $\frac{1}{2}$	&   $\frac{1}{2}^-$  &	4071&4048&4074&3998&&4136\\
			$\Xi_{cc} (\frac{3}{2}^-, \frac{1}{2})$&	0	&	0	&	0	&	1	&	1   &   1  &  $\frac{1}{2}$	&   $\frac{3}{2}^-$ & 	4085&4081&4078&4014&&4196\\
			$\Xi_{cc} (\frac{1}{2}^-, \frac{3}{2})$&    0	&	0	&	0	&	1	&	1   &   1  &  $\frac{3}{2}$	&	$\frac{1}{2}^-$ & 	4073&4082&&3985&&4053 \\
			$\Xi_{cc} (\frac{3}{2}^-, \frac{3}{2})$&	0	&	0	&	0	&	1	&	1   &	1  &  $\frac{3}{2}$	&   $\frac{3}{2}^-$ & 	4095&4114 &&4025&&4101 \\
			$\Xi_{cc} (\frac{5}{2}^-, \frac{3}{2})$&	0	&	0	&	0	&	1	&	1   &	1  &  $\frac{3}{2}$	&	$\frac{5}{2}^-$ &	4099&4169 &4092&4050&&4155 \\
   	\hline	\hline
    		States& $n_{\rho}$ & $n_{\lambda}$ & $l_{\rho}$ & $l_{\lambda}$ & $S_{\rho}$ & $J_{\rho}$ & $j$	&  $J^P$&Mass&\cite{Ortiz-Pacheco:2023kjn}&\cite{Roberts:2007ni}&\cite{Lu:2017meb}&\cite{Yu:2022lel}&\cite{Ebert:2002ig}\\
			\hline
    $\Omega_{cc}(1S)$	  &	0	&	0	&	0	&	0	&	1   &1    &$\frac{1}{2}$	&	$\frac{1}{2}^+$  & 3768&3766 &3815&3715&3750&3778\\
			$\Omega_{cc}^{*}(1S)$	  &	0	&	0	&	0	&	0	&	1   &1    &$\frac{1}{2}$	&	$\frac{3}{2}^+$  & 3819& 3833& 3876&3772&3799&3872\\ 
  	$\tilde{\Omega}_{cc}(\frac{1}{2}^-, \frac{1}{2})$	  &	0	&	0	&	1	&	0	&	0   &1    &$\frac{1}{2}$	&	$\frac{1}{2}^-$  &4022 & 3970&4046&3986&4049 & 4002\\
			$\tilde{\Omega}_{cc}(\frac{3}{2}^-, \frac{1}{2})$	  &	0	&	0	&	1	&	0	&	0   &1    &$\frac{1}{2}$	&	$\frac{3}{2}^-$  &4022 &4002 &4052&4020&4089 & 4102\\
			$\Omega_{cc}(\frac{1}{2}^-, \frac{1}{2})$	  &	0	&	0	&	0	&	1	&	1   &1    &$\frac{1}{2}$	&	$\frac{1}{2}^-$  &4135 &4111 &4135&4087&&4271\\
			$\Omega_{cc}(\frac{3}{2}^-, \frac{1}{2})$	  &	0	&	0	&	0	&	1	&	1   &1    &$\frac{1}{2}$	&	$\frac{3}{2}^-$   &4146 &4144 &4140&4107&&4325\\
			$\Omega_{cc}(\frac{1}{2}^-, \frac{3}{2})$	  &	0	&	0	&	0	&	1	&	1   &1    &$\frac{3}{2}$	&	$\frac{1}{2}^-$  &4137 &4145 &&4081&&4208\\
			$\Omega_{cc}(\frac{3}{2}^-, \frac{3}{2})$	  &	0	&	0	&	0	&	1 &	1   &1    &$\frac{3}{2}$	&	$\frac{3}{2}^-$   &4154 &4177 &&4114&&4252\\
			$\Omega_{cc}(\frac{5}{2}^-, \frac{3}{2})$	  &	0	&	0	&	0	&	1	&	1   &1    &$\frac{3}{2}$	&	$\frac{5}{2}^-$   &4156 &4232 &4152&4134&&4303\\
   	\hline\hline
		\end{tabular*}
	\end{center}
\end{table*}

\begin{table*}[!htbp]
	\begin{center}
		\caption{\label{mass2} The mass spectra of $\Xi_{bb}$ and $\Omega_{bb}$ in MeV.}
		\renewcommand{\arraystretch}{1.5}
		\normalsize
		\begin{tabular*}{18cm}{@{\extracolsep{\fill}}p{1.9cm}<{\centering}p{0.2cm}<{\centering}p{0.2cm}<{\centering}p{0.2cm}<{\centering}p{0.2cm}<{\centering}p{0.2cm}<{\centering}p{0.2cm}<{\centering}p{0.2cm}<{\centering}p{0.2cm}<{\centering}p{0.8cm}<{\centering}p{0.8cm}<{\centering}p{0.8cm}<{\centering}p{0.8cm}<{\centering}p{0.8cm}<{\centering}p{0.8cm}<{\centering}}
			\hline\hline
			States& $n_{\rho}$ & $n_{\lambda}$ & $l_{\rho}$ & $l_{\lambda}$ & $S_{\rho}$ & $J_{\rho}$ & $j$	&  $J^P$&Mass&\cite{Ortiz-Pacheco:2023kjn}&\cite{Roberts:2007ni}&\cite{Lu:2017meb}&\cite{Li:2022oth}&\cite{Ebert:2002ig}\\
			\hline
			$\Xi_{bb} (1S)$	  &	0	&	0	&	0	&	0	&	1   &1    &$\frac{1}{2}$	&	$\frac{1}{2}^+$  & 10250&10295 &10340&10138&10192&10202\\
			$\Xi_{bb}^{*}(1S)$      &	0	&	0	&	0	&	0	&	1   &1    &$\frac{1}{2}$	&	$\frac{3}{2}^+$   & 10275&10317&10367&10169&10211 &10237\\ 
   $\tilde{\Xi}_{bb} (\frac{1}{2}^-, \frac{1}{2})$&	0	&	0	&	1	&	0	&	0   &   1  &  $\frac{1}{2}$	&   $\frac{1}{2}^-$  &	10412&10411&10493&10364&10428 & 10368\\
			$\tilde{\Xi}_{bb} (\frac{3}{2}^-, \frac{1}{2})$&	0	&	0	&	1	&	0	&	0   &   1  &  $\frac{1}{2}$	&   $\frac{3}{2}^-$  &	10412&10417&10495&10387&10445 & 10408\\
			$\Xi_{bb} (\frac{1}{2}^-, \frac{1}{2})$&	0	&	0	&	0	&	1	&	1   &   1  &  $\frac{1}{2}$	&   $\frac{1}{2}^-$  &	10639&10700&10710&10525&&10675\\
			$\Xi_{bb} (\frac{3}{2}^-, \frac{1}{2})$&	0	&	0	&	0	&	1	&	1   &   1  &  $\frac{1}{2}$	&   $\frac{3}{2}^-$  &	10676&10707&10713&10526&&10694\\
			$\Xi_{bb} (\frac{1}{2}^-, \frac{3}{2})$&    0	&	0	&	0	&	1	&	1   &   1  &  $\frac{3}{2}$	&	$\frac{1}{2}^-$  &	10640&10716&&10504&&10632 \\
			$\Xi_{bb} (\frac{3}{2}^-, \frac{3}{2})$&	0	&	0	&	0	&	1	&	1   &	1  &  $\frac{3}{2}$	&   $\frac{3}{2}^-$  &	10678&10722&&10528&&10647\\
			$\Xi_{bb} (\frac{5}{2}^-, \frac{3}{2})$&	0	&	0	&	0	&	1	&	1   &	1  &  $\frac{3}{2}$	&	$\frac{5}{2}^-$ &	10695&10733&10713&10547&&10661 \\
   	\hline\hline
    	States& $n_{\rho}$ & $n_{\lambda}$ & $l_{\rho}$ & $l_{\lambda}$ & $S_{\rho}$ & $J_{\rho}$ & $j$	&  $J^P$&Mass&\cite{Ortiz-Pacheco:2023kjn}&\cite{Roberts:2007ni}&\cite{Lu:2017meb}&\cite{Li:2022oth}&\cite{Ebert:2002ig}\\\hline
    $\Omega_{bb}(1S)$	  &	0	&	0	&	0	&	0	&	1   &1    &$\frac{1}{2}$	&	$\frac{1}{2}^+$  & 10383&10438 &10454&10230&10285&10359\\
			$\Omega_{bb}^{*}(1S)$	  &	0	&	0	&	0	&	0	&	1   &1    &$\frac{1}{2}$	&	$\frac{3}{2}^+$  & 10403& 10460&10486&10258&10303&10389 \\ 
  	$\tilde{\Omega}_{bb}(\frac{1}{2}^-, \frac{1}{2})$	  &	0	&	0	&	1	&	0	&	0   &1    &$\frac{1}{2}$	&	$\frac{1}{2}^-$ &10543&10554&10616&10464 &10528 & 10532\\
			$\tilde{\Omega}_{bb}(\frac{3}{2}^-, \frac{1}{2})$	  &	0	&	0	&	1	&	0	&	0   &1    &$\frac{1}{2}$	&	$\frac{3}{2}^-$  &10544 &10560&10619&10482&10543 & 10566\\
			$\Omega_{bb}(\frac{1}{2}^-, \frac{1}{2})$	  &	0	&	0	&	0	&	1	&	1   &1    &$\frac{1}{2}$	&	$\frac{1}{2}^-$ &  10732&10762&10763&10605&&10804\\
			$\Omega_{bb}(\frac{3}{2}^-, \frac{1}{2})$	  &	0	&	0	&	0	&	1	&	1   &1    &$\frac{1}{2}$	&	$\frac{3}{2}^-$  &10739 &10768&10765&10610&&10821\\
			$\Omega_{bb}(\frac{1}{2}^-, \frac{3}{2})$	  &	0	&	0	&	0	&	1	&	1   &1    &$\frac{3}{2}$	&	$\frac{1}{2}^-$ &10733 &10778 & &10591&&10771\\
			$\Omega_{bb}(\frac{3}{2}^-, \frac{3}{2})$	  &	0	&	0	&	0	&	1	&	1   &1    &$\frac{3}{2}$	&	$\frac{3}{2}^-$  &10741 &10784 &&10611&&10785\\
			$\Omega_{bb}(\frac{5}{2}^-, \frac{3}{2})$	  &	0	&	0	&	0	&	1	&	1   &1    &$\frac{3}{2}$	&	$\frac{5}{2}^-$ &10744 &10795 & 10766&10625&&10798\\
   	\hline\hline
		\end{tabular*}
	\end{center}
\end{table*}

The helicity amplitude $\mathcal{A}_{h}$ can be derived from the transition operator and the initial and final wave functions. In order to calculate the strong decay widths for pseudoscalar meson emissions, the phase space factor also needs to be taken into account. Then, the strong decays for doubly charmed and bottom baryons can be evaluated straightforwardly
\begin{equation}{\label{decay}}
\Gamma=\frac{1}{4 \pi} \frac{|\boldsymbol{q}|}{2 M_{i}^{2}} \frac{1}{2 J+1} \sum_{h}\left|\mathcal{A}_{h}\right|^{2}.
\end{equation}

\section{Results and discussions}{\label{low-lying}}

\subsection{The $\lambda-$mode  $\Xi_{cc}(1P)$  and $\Omega_{cc}(1P)$ states}

There are five $\lambda-$mode $\Xi_{cc} (1P)$ states in the constituent quark model, which can be classified into the  $j=1/2$ doublet and $j=3/2$ triplet according to the light quark spin $j$. The strong decays for $\lambda-$mode $\Xi_{cc}(1P)$ states are listed in Table~\ref{1pcc}. For the $j=1/2$ doublet, the predicted decay widths are broad, which are about 297 and 272 MeV, respectively. Also, it can be found that the strong decays for $\Xi_{cc}(1/2^{-},1/2)$ and $\Xi_{cc}(3/2^{-},1/2)$ states are dominated by the $\Xi_{cc}\pi$ and $\Xi_{cc}^{*}\pi$ channels, respectively. However, for the $j=3/2$ triplet,  the predicted decay widths are relatively narrow and about 46, 67, and 82 MeV for $J^{PC}=1/2^{-}$, $3/2^{-}$, and $5/2^{-}$ states respectively. The $\Xi_{cc}^*\pi$ channel saturates the strong decay of $\Xi_{cc}(1/2^{-},3/2)$ state, and the branching ratios for $\Xi_{cc}(3/2^{-},3/2)$ and $\Xi_{cc}(5/2^{-},3/2)$ states are calculated to be 
\begin{equation}
    Br(\Xi_{cc}\pi,\Xi_{cc}^{*}\pi)=48.47\%,51.53\%,
\end{equation}
and 
\begin{equation}
    Br(\Xi_{cc}\pi,\Xi_{cc}^{*}\pi)=45.55\%,54.45\%,
\end{equation}
respectively. 

\begin{table}[htbp]
	\caption{\label{1pcc}The predicted strong decay widths for $\lambda-$mode $\Xi_{cc}(1P)$ and $\Omega_{cc} (1P)$ states in MeV. The $\times$ denotes the forbidden channel due to quantum numbers.}
	\renewcommand{\arraystretch}{1.5}
	\begin{ruledtabular}
		\begin{tabular}{ccccccc}
			State
			&$\Xi_{cc}(\frac{1}{2}^{-},\frac{1}{2})$
			&$\Xi_{cc}(\frac{3}{2}^{-},\frac{1}{2})$
			&$\Xi_{cc}(\frac{1}{2}^{-},\frac{3}{2})$
			&$\Xi_{cc}(\frac{3}{2}^{-},\frac{3}{2})$
			&$\Xi_{cc}(\frac{5}{2}^{-},\frac{3}{2})$
			\\\hline
			$\Xi_{cc}$ $\pi$&296.88&×&×&19.62&53.52\\
			$\Xi_{cc}^{*}$ $\pi$&×&271.59&46.38&46.92&28.53\\
			Total &296.88&271.59&46.38&66.54&82.05\\
			\hline\hline
			State
			&$\Omega_{cc}(\frac{1}{2}^{-},\frac{1}{2})$
			&$\Omega_{cc}(\frac{3}{2}^{-},\frac{1}{2})$
			&$\Omega_{cc}(\frac{1}{2}^{-},\frac{3}{2})$
			&$\Omega_{cc}(\frac{3}{2}^{-},\frac{3}{2})$
			&$\Omega_{cc}(\frac{5}{2}^{-},\frac{3}{2})$
			\\\hline
			$\Xi_{cc}$ $\bar{K}$&349.23&×&×&0.39&1.20\\
            
			Total &349.23&Narrow&Narrow&0.39&1.20\\
		\end{tabular}
	\end{ruledtabular}
\end{table}

The broad $j = 1/2$ doublet and narrow $j = 3/2$ triplet result from the enhancement or cancellation in the amplitudes with different Clebsch-Gordan coefficients. In the heavy superflavor symmetry, the $cc$ subsystem corresponds to an antiquark $\bar c$, and the $\lambda-$mode $\Xi_{cc}(1P)$ states are related to $P-$wave charmed mesons. Indeed, from the Review of Particle Physics~\cite{ParticleDataGroup:2024cfk}, the $D_0^*(2300)$ and $D_1(2430)$ with $j=1/2$ are broad resonances, and the $D_1(2420)$ and $D_2(2460)$ with $j=3/2$ are relatively narrow. It can be seen that the heavy superflavor symmetry is preserved well for decay behaviors of $\Xi_{cc}(1P)$ baryons and $D(1P)$ mesons. 

According to the SU(3) flavor symmetry of light quarks, the $\Omega_{cc}(1P)$ states should show similar properties to the $\Xi_{cc}(1P)$ states. However, in present calculations, only the $\Omega_{cc}(1/2^-,1/2)$ state has a broad width that decays into the $\Xi_{cc}\bar K$ channel, while other four states are predicted to be rather narrow. The main reason for this broken symmetry is that some $\lambda-$mode $\Omega_{cc}(1P)$ states lie near the $\Xi_{cc} \bar K$ threshold and below the $\Xi_{cc}^* \bar K$ threshold, which leads to the narrow widths for the $\Omega_{cc}(3/2^-,1/2)$, $\Omega_{cc}(1/2^-,3/2)$, $\Omega_{cc}(3/2^-,3/2)$, and $\Omega_{cc}(5/2^-,3/2)$ states. Then, the isospin broken pion emissions and radiative transitions become important and may dominate the decay behaviors of $\lambda-$mode $\Omega_{cc}(1P)$ states except for the $\Omega_{cc}(1/2^-,1/2)$ state. The similar situation occurs in the charmed-strange mesons, where the particles $D_{s0}^*(2137)$ and $D_{s1}(2460)$ have significantly lower masses and narrow decay widths. For the $D_{s0}^*(2137)$ and $D_{s1}(2460)$ resonances, the molecular interpretation and hybrid picture with coupled channel approach were proposed in the literature to explain their mysterious properties~\cite{Hwang:2004cd,MartinezTorres:2011pr,Mohler:2013rwa,Lang:2014yfa,Ortega:2016mms,Bali:2017pdv,Albaladejo:2018mhb}. Also, the $\lambda-$mode $\Omega_{cc}(1P)$ states may couple to the meson-baryon molecular states~\cite{Yan:2018zdt,Wang:2021rjk,Wang:2022aga,Wang:2023mdj}, which lower their masses of three-quark pictures. In brief, both our calculations and heavy superflavor symmetry indicate that there exist several negative-parity $\Omega_{cc}(1P)$ states near the $\Xi_{cc} \bar K$ threshold and below the $\Xi_{cc}^* \bar K$ threshold. We suggest that the future experiments search for these states through isospin broken decay modes $\Omega_{cc}^{(*)}\pi$ and radiative decays $\Omega_{cc}^{(*)}\gamma$, which are also useful for us to better understand the nature of negative-parity charmed-strange mesons. 
 
\subsection{The $\lambda-$mode $\Xi_{bb}(1P)$  and $\Omega_{bb}(1P)$ states}

The  strong decays for $\lambda-$mode $\Xi_{bb}(1P)$ and $\Omega_{bb}(1P)$ states are estimated and shown in Table~\ref{bb1p}. For the $j=1/2$ doublet, the calculated decay widths are rather broad with 303 and 312 MeV. For the $j=3/2$ triplet, the predicted decay widths are about 41, 67, and 84 MeV for the $\Xi_{bb}(1/2^{-},3/2)$, $\Xi_{bb}(3/2^{-},3/2)$ and $\Xi_{bb}(5/2^{-},3/2)$ states, respectively. Meanwhile, the total decay widths for the five $\lambda-$mode $\Omega_{bb}(1P)$ states are rather narrow owing to the lower masses and closed $\Xi_{bb}\bar K$ channel. These narrow states can be hunted for through the isospin broken pion emissions and radiative transitions in future experiments. Moreover, the narrow $\Omega_{bb}(1P)$ states may be observed earlier than the ground state $\Omega_{bb}^*(1S)$ in the future as well as the singly bottom $\Omega_{b}$ family~\cite{ParticleDataGroup:2024cfk}. 

\begin{table}[htbp]
	\caption{\label{bb1p}The predicted strong decay widths for $\lambda-$mode $\Xi_{bb}(1P)$ and $\Omega_{bb} (1P)$ states in MeV. The $\times$ denotes the forbidden channel due to quantum numbers, and the $\cdot \cdot \cdot$ stands for the the forbidden channel owing to the lack of phase space.}
	\renewcommand{\arraystretch}{1.5}
	\begin{ruledtabular}
		\begin{tabular}{ccccccc}
			State
			&$\Xi_{bb}(\frac{1}{2}^{-},\frac{1}{2})$
			&$\Xi_{bb}(\frac{3}{2}^{-},\frac{1}{2})$
			&$\Xi_{bb}(\frac{1}{2}^{-},\frac{3}{2})$
			&$\Xi_{bb}(\frac{3}{2}^{-},\frac{3}{2})$
			&$\Xi_{bb}(\frac{5}{2}^{-},\frac{3}{2})$
			\\\hline
			$\Xi_{bb}$ $\pi$&302.94&×&×&15.81&48.48\\
			$\Xi_{bb}^{*}$ $\pi$&×&311.22&40.89&51.57&35.37\\
			Total &302.94&311.22&40.89&67.38&83.85
            \\
			\hline\hline
			State
			&$\Omega_{bb}(\frac{1}{2}^{-},\frac{1}{2})$
			&$\Omega_{bb}(\frac{3}{2}^{-},\frac{1}{2})$
			&$\Omega_{bb}(\frac{1}{2}^{-},\frac{3}{2})$
			&$\Omega_{bb}(\frac{3}{2}^{-},\frac{3}{2})$
			&$\Omega_{bb}(\frac{5}{2}^{-},\frac{3}{2})$
			\\\hline
			$\Xi_{bb}$ $\bar{K}$&$\cdot\cdot\cdot$&$\cdot\cdot\cdot$$($×$)$&$\cdot\cdot\cdot$$($×$)$&$\cdot\cdot\cdot$&0.00\\
            	
    			Total &Narrow&Narrow&Narrow&Narrow&Narrow\\
		\end{tabular}
	\end{ruledtabular}
\end{table}

Our results suggest that the strong decay behaviors for $\lambda-$mode $\Xi_{bb} (1P)$ and $\Omega_{bb}(1P)$ states are quite similar to that of charmed sectors. Therefore, the correspondence between the $cc$ and $bb$ subsystems preserves well for these states. Also, the approximate superflavor symmetry relates the $bb$ subsystem to an antiquark $\bar b$ and the doubly bottom baryons to the bottom(-strange) mesons. For the bottom(-strange) mesons, four narrow states $B_1(5721)$, $B_2^*(5747)$, $B_{s1}(5830)$, and $B_{s2}^*(5840)$ are observed experimentally, which just occupy the four $P-$wave $j=3/2$ states in the quark model~\cite{Lu:2016bbk}. Meanwhile, no experimental evidence for the four $j=1/2$ bottom(-strange) mesons exists. More theoretical and experimental information on the bottom(-strange) mesons is also helpful for understanding the doubly bottom baryons.


\subsection{Further discussions}

We first compare our present results of $\lambda-$mode doubly heavy charmed and bottomed baryons with other previous calculations within the quark models in the literature. For the $\Xi_{cc/bb}(1P)$ states, the broad $j = 1/2$ doublet and narrow $j = 3/2$ triplet are key characteristics in present results, which roughly agree with the $^3P_0$ model calculations for the $\Xi_{cc}(1P)$~\cite{Eakins:2012fq} and $\Xi_{bb}(1P)$~\cite{He:2021iwx,Chen:2022fye}. For the $\Omega_{cc}(1P)$ states, there is no other work for comparison. The strong decay behaviors for $\Omega_{bb}(1P)$ show different properties with those of $^3P_0$ model calculations~\cite{He:2021iwx,Chen:2022fye}, where the variant phase spaces can affect the decay behaviors significantly for the near threshold states. It should be noted that the calculations in Refs.~\cite{Xiao:2017udy,Xiao:2017dly} are based on the $L-S$ coupling scheme for the doubly heavy baryons, which cannot be compared with present calculations directly. Further careful theoretical investigations and experimental inputs are helpful to understand their decay behaviors.      

Besides the $\lambda-$mode excitations, there are also $\rho-$mode $1P$ states denoted as $\tilde{\Xi}_{cc/bb}(1P)/\tilde{\Omega}_{cc/bb}(1P)$ states, and their masses are listed in Table~\ref{mass1} and~\ref{mass2}. Owing to the limited phase space, the only possible strong decay modes are light meson emissions for these low-lying states. However, under the spectator assumption for the two heavy quark subsystems, the orthogonality of $\rho-$mode wave functions between initial and final states leads to the vanishing amplitude in the tree level diagram shown in Figure~\ref{emission}. Therefore, the light meson emissions for these low-lying $\rho-$mode states are highly suppressed and  the electromagnetic and weak decays may become dominating. More discussions can be found in Refs.~\cite{Eakins:2012fq,Song:2023cyk}.

From our calculations on masses and strong decays for doubly charmed and bottom baryons, it can be seen that the heavy superflavor symmetry is preserved well in most cases. Although the masses of charm and bottom quarks are finite, the violation of superflavor symmetry owing to the heavy quark masses and different phase spaces is quite small. This is consistent with the the previous discussions in the quark model~\cite{Eakins:2012jk,Eakins:2012fq,He:2021iwx,Chen:2022fye} and effective field theory~\cite{Yan:2018zdt,Shi:2020qde}, but differ from the results of Ref.~\cite{Soto:2020pfa} in which the next leading order corrections can be considered as an important source for the violation. In fact, our decay calculations are based on a spectator model, where the two heavy quarks as a whole in the initial baryons go into the final states and the heavy superflavor symmetry for the decay amplitude is preserved automatically. 

\section{SUMMARY}{\label{SUMMARY}}

 In this work, we investigate the strong decays for low-lying excited states of doubly charmed and bottom baryons in the constituent quark model. For the $\lambda-$mode $\Xi_{cc/bb}(1P)$ states, the $j=1/2$ doublet are rather broad and the $j=3/2$ triplet are expected to be relatively narrow. For the $\lambda-$mode $\Omega_{cc/bb}(1P)$ states, only the $\Omega_{cc}(1/2^-,1/2)$ state is broad, while other states are rather narrow owing to the limited phase space and selection rule. Moreover,  the light meson emissions for the low-lying $\rho-$mode states are highly suppressed due to the orthogonality of wave functions between initial and final states, and then the electromagnetic and weak decay channels for these states may become dominating.

Also, the strong decay behaviors for doubly charmed and bottom baryons preserve the heavy superflavor symmetry well, where the small violation originates from the finite heavy quark masses and different phase spaces. Our results indicate that one can establish a close connection between doubly heavy baryons and heavy mesons and safely apply this corresponding relationship to study the two kinds of systems simultaneously. We hope these theoretical results of masses and strong decays for undiscovered doubly charmed and bottom baryons can provide helpful information for future experiments and help us to better understand the heavy quark symmetry.

\section*{ACKNOWLEDGMENTS}
This work is supported by the Natural Science Foundation of  Hunan Province under Grant No. 2023JJ40421, the
Scientific Research Foundation of Hunan Provincial Education Department under Grant No. 24B0063, and the Youth Talent
Support Program of Hunan Normal University under Grant
No. 2024QNTJ14.


\begin{thebibliography}{99}
	
	
\bibitem{ParticleDataGroup:2024cfk}
S.~Navas \textit{et al.} (Particle Data Group),
Review of particle physics,
Phys. Rev. D \textbf{110}, 030001 (2024).



\bibitem{Mattson:2002vu}
M.~Mattson {\it et al.} (SELEX Collaboration),
First observation of the doubly charmed baryon $\Xi_{cc}^+$,
Phys.\ Rev.\ Lett.\  {\bf 89}, 112001 (2002).

\bibitem{Ocherashvili:2004hi}
A.~Ocherashvili {\it et al.} (SELEX Collaboration),
Confirmation of the double charm baryon $\Xi^+_{cc}(3520)$ via its decay to $pD^+K^-$,
Phys.\ Lett.\ B {\bf 628}, 18 (2005).

\bibitem{Ratti:2003ez}
S.~P.~Ratti,
New results on $c$-baryons and a search for $cc$-baryons in FOCUS,
Nucl.\ Phys.\ Proc.\ Suppl.\  {\bf 115}, 33 (2003).

\bibitem{Aubert:2006qw}
B.~Aubert {\it et al.} (BaBar Collaboration),
Search for doubly charmed baryons $\Xi_{cc}^+$ and $\Xi_{cc}^{++}$ in BABAR,
Phys.\ Rev.\ D {\bf 74}, 011103 (2006).

\bibitem{Chistov:2006zj}
R.~Chistov {\it et al.} (Belle Collaboration),
Observation of new states decaying into $\Lambda_c^+ K^- \pi^+$ and $\Lambda_c^+ K^0_S \pi^-$,
Phys.\ Rev.\ Lett.\  {\bf 97}, 162001 (2006).

\bibitem{Aaij:2013voa}
R.~Aaij {\it et al.} (LHCb Collaboration),
Search for the doubly charmed baryon $\Xi_{cc}^+$,
JHEP {\bf 1312}, 090 (2013).

\bibitem{Aaij:2017ueg}
R.~Aaij \textit{et al.} (LHCb Collaboration),
Observation of the doubly charmed baryon $\Xi_{cc}^{++}$,
Phys. Rev. Lett. \textbf{119}, 112001 (2017).

\bibitem{LHCb:2018zpl}
R.~Aaij \textit{et al.} (LHCb Collaboration),
Measurement of the Lifetime of the Doubly Charmed Baryon $\Xi_{cc}^{++}$,
Phys. Rev. Lett. \textbf{121}, 052002 (2018).

\bibitem{LHCb:2019qed}
R.~Aaij \textit{et al.} (LHCb Collaboration),
Measurement of $\mathit{\Xi}_{cc}^{++}$ production in $pp$ collisions at $\sqrt{s}=13$ TeV,
Chin. Phys. C \textbf{44}, 022001 (2020).


\bibitem{LHCb:2019gqy}
R.~Aaij \textit{et al.} (LHCb Collaboration),
Search for the doubly charmed baryon $\Xi_{cc}^+$,
Sci. China Phys. Mech. Astron. \textbf{63}, 221062 (2020).

%
\bibitem{LHCb:2021eaf}
R.~Aaij \textit{et al.} (LHCb Collaboration),
Search for the doubly charmed baryon $ {\varXi}_{cc}^{+} $ in the $ {\varXi}_c^{+}{\pi}^{-}{\pi}^{+} $ final state,
JHEP \textbf{12}, 107 (2021).

\bibitem{LHCb:2021rkb}
R.~Aaij \textit{et al.} (LHCb Collaboration),
Search for the doubly charmed baryon \ensuremath{\Omega}$_{cc}^{+}$,
Sci. China Phys. Mech. Astron. \textbf{64}, 101062 (2021).

\bibitem{LHCb:2022fbu}
R.~Aaij \textit{et al.} (LHCb Collaboration),
Search for the doubly heavy baryon $\it{\Xi}_{bc}^{+}$ decaying to $J/\it{\psi} \it{\Xi}_{c}^{+}$,
Chin. Phys. C \textbf{47}, 093001 (2023).


\bibitem{LHCb:2020iko}
R.~Aaij \textit{et al.} (LHCb Collaboration),
Search for the doubly heavy $\Xi_{bc}^0$ baryon via decays to $D^0pK^-$,
JHEP \textbf{11}, 095 (2020).


\bibitem{LHCb:2021xba}
R.~Aaij \textit{et al.} (LHCb Collaboration),
Search for the doubly heavy baryons $\Omega^0_{bc}$ and $\Xi^0_{bc}$ decaying to $\Lambda^+_c \pi^-$ and $\Xi^+_c \pi^-$,
Chin. Phys. C \textbf{45}, 093002 (2021).


\bibitem{LHCb:2021vvq}
R.~Aaij \textit{et al.} (LHCb Collaboration),
Observation of an exotic narrow doubly charmed tetraquark,
Nature Phys. \textbf{18}, 751-754 (2022).

\bibitem{LHCb:2021auc}
R.~Aaij \textit{et al.} (LHCb Collaboration),
Study of the doubly charmed tetraquark $T_{cc}^{+}$,
Nature Commun. \textbf{13}, 3351 (2022).



\bibitem{Ebert:1996ec}
D.~Ebert, R.~N.~Faustov, V.~O.~Galkin, A.~P.~Martynenko and V.~A.~Saleev,
Heavy baryons in the relativistic quark model,
Z.\ Phys.\ C {\bf 76}, 111 (1997).
\bibitem{Kiselev:2001fw}
V.~V.~Kiselev and A.~K.~Likhoded,
Baryons with two heavy quarks,
Phys.\ Usp.\  {\bf 45}, 455 (2002).



\bibitem{Gershtein:2000nx}
S.~S.~Gershtein, V.~V.~Kiselev, A.~K.~Likhoded and A.~I.~Onishchenko,
Spectroscopy of doubly heavy baryons,
Phys.\ Rev.\ D {\bf 62}, 054021 (2000).


\bibitem{Tong:1999qs}
S.~P.~Tong, Y.~B.~Ding, X.~H.~Guo, H.~Y.~Jin, X.~Q.~Li, P.~N.~Shen and R.~Zhang,
Spectra of baryons containing two heavy quarks in potential model,
Phys. Rev. D \textbf{62}, 054024 (2000).

\bibitem{Ortiz-Pacheco:2023kjn}
E.~Ortiz-Pacheco and R.~Bijker,
Masses and radiative decay widths of S- and P-wave singly, doubly, and triply heavy charm and bottom baryons,
Phys. Rev. D \textbf{108}, 054014 (2023).


\bibitem{Roberts:2007ni}
W.~Roberts and M.~Pervin,
Heavy baryons in a quark model,
Int.\ J.\ Mod.\ Phys.\ A {\bf 23}, 2817 (2008).

\bibitem{Lu:2017meb}
Q.~F.~L\"u, K.~L.~Wang, L.~Y.~Xiao and X.~H.~Zhong,
Mass spectra and radiative transitions of doubly heavy baryons in a relativized quark model,
Phys. Rev. D \textbf{96}, 114006 (2017).


\bibitem{Yu:2022lel}
G.~L.~Yu, Z.~Y.~Li, Z.~G.~Wang, J.~Lu and M.~Yan,
Systematic analysis of doubly charmed baryons $\Xi _{cc}$ and $\Omega _{cc}$,
Eur. Phys. J. A \textbf{59}, 126 (2023).



\bibitem{Li:2022oth}
Z.~Y.~Li, G.~L.~Yu, Z.~G.~Wang, J.~Z.~Gu and H.~T.~Shen,
Mass spectra of double-bottom baryons,
Mod. Phys. Lett. A \textbf{38}, 2350052 (2023).

\bibitem{Ebert:2002ig}
D.~Ebert, R.~N.~Faustov, V.~O.~Galkin and A.~P.~Martynenko,
Mass spectra of doubly heavy baryons in the relativistic quark model,
Phys.\ Rev.\ D {\bf 66}, 014008 (2002).

\bibitem{Valcarce:2008dr}
A.~Valcarce, H.~Garcilazo and J.~Vijande,
Towards an understanding of heavy baryon spectroscopy,
Eur.\ Phys.\ J.\ A {\bf 37}, 217 (2008).


\bibitem{Eakins:2012jk}
B.~Eakins and W.~Roberts,
Symmetries and Systematics of Doubly Heavy Hadrons,
Int. J. Mod. Phys. A \textbf{27}, 1250039 (2012).


\bibitem{Yoshida:2015tia}
T.~Yoshida, E.~Hiyama, A.~Hosaka, M.~Oka and K.~Sadato,
Spectrum of heavy baryons in the quark model,
Phys. Rev. D \textbf{92},  114029 (2015).

\bibitem{Shah:2017liu}
Z.~Shah and A.~K.~Rai,
Excited state mass spectra of doubly heavy $\Xi$ baryons,
Eur.\ Phys.\ J.\ C {\bf 77}, 129 (2017).

\bibitem{Silvestre-Brac:1996myf}
B.~Silvestre-Brac,
Spectrum and static properties of heavy baryons,
Few Body Syst. \textbf{20}, 1-25 (1996).

\bibitem{Soto:2020pfa}
J.~Soto and J.~Tarr\'us Castell\`a,
Effective field theory for double heavy baryons at strong coupling,
Phys. Rev. D \textbf{102}, 014013 (2020); erratum: Phys. Rev. D \textbf{104}, 059901 (2021).

\bibitem{Savage:1990di}
M.~J.~Savage and M.~B.~Wise,
Spectrum of baryons with two heavy quarks,
Phys. Lett. B \textbf{248}, 177-180 (1990).

\bibitem{Cohen:2006jg}
T.~D.~Cohen and P.~M.~Hohler,
Doubly heavy hadrons and the domain of validity of doubly heavy diquark-anti-quark symmetry,
Phys.\ Rev.\ D {\bf 74}, 094003 (2006).


\bibitem{Ma:2022vqf}
Y.~Ma, L.~Meng, Y.~K.~Chen and S.~L.~Zhu,
Ground state baryons in the flux-tube three-body confinement model using diffusion Monte~Carlo,
Phys. Rev. D \textbf{107}, 054035 (2023).

\bibitem{Wei:2015gsa}
K.~W.~Wei, B.~Chen and X.~H.~Guo,
Masses of doubly and triply charmed baryons,
Phys.\ Rev.\ D {\bf 92}, 076008 (2015).

\bibitem{Song:2022csw}
Y.~Song, D.~Jia, W.~Zhang and A.~Hosaka,
Low-lying doubly heavy baryons: Regge relation and mass scaling,
Eur. Phys. J. C \textbf{83}, 1 (2023).

\bibitem{Aliev:2012nn}
T.~M.~Aliev, K.~Azizi and M.~Savci,
Mixing angle of doubly heavy baryons in QCD,
Phys.\ Lett.\ B {\bf 715}, 149 (2012).


\bibitem{Liu:2009jc}
L.~Liu, H.~W.~Lin, K.~Orginos and A.~Walker-Loud,
Singly and Doubly Charmed J=1/2 Baryon Spectrum from Lattice QCD,
Phys.\ Rev.\ D {\bf 81}, 094505 (2010).

\bibitem{Brown:2014ena}
Z.~S.~Brown, W.~Detmold, S.~Meinel and K.~Orginos,
Charmed bottom baryon spectroscopy from lattice QCD,
Phys.\ Rev.\ D {\bf 90}, 094507 (2014).

\bibitem{Padmanath:2015jea}
M.~Padmanath, R.~G.~Edwards, N.~Mathur and M.~Peardon,
Spectroscopy of doubly-charmed baryons from lattice QCD,
Phys.\ Rev.\ D {\bf 91}, 094502 (2015).

\bibitem{Mathur:2018rwu}
N.~Mathur and M.~Padmanath,
Lattice QCD study of doubly-charmed strange baryons,
Phys. Rev. D \textbf{99}, 031501 (2019).


\bibitem{Mathur:2018epb}
N.~Mathur, M.~Padmanath and S.~Mondal,
Precise predictions of charmed-bottom hadrons from lattice QCD,
Phys. Rev. Lett. \textbf{121}, 202002 (2018).

\bibitem{Albertus:2009ww}
C.~Albertus, E.~Hernandez and J.~Nieves,
Hyperfine mixing in $b \to c$ semileptonic decay of doubly heavy baryons,
Phys.\ Lett.\ B {\bf 683}, 21 (2010).

\bibitem{White:1991hz}
M.~J.~White and M.~J.~Savage,
Semileptonic decay of baryons with two heavy quarks,
Phys.\ Lett.\ B {\bf 271}, 410 (1991).

\bibitem{Li:2017ndo}
R.~H.~Li, C.~D.~L\"{u}, W.~Wang, F.~S.~Yu and Z.~T.~Zou,
Doubly-heavy baryon weak decays: $\Xi_{bc}^{0}\to pK^{-}$ and $\Xi_{cc}^{+}\to \Sigma_{c}^{++}(2520)K^{-}$,
Phys.\ Lett.\ B {\bf 767}, 232 (2017).

\bibitem{Yu:2017zst}
F.~S.~Yu, H.~Y.~Jiang, R.~H.~Li, C.~D.~L\"u, W.~Wang and Z.~X.~Zhao,
Discovery Potentials of Doubly Charmed Baryons,
Chin. Phys. C \textbf{42}, 051001 (2018).

\bibitem{Chen:2016spr}
H.~X.~Chen, W.~Chen, X.~Liu, Y.~R.~Liu and S.~L.~Zhu,
A review of the open charm and open bottom systems,
Rept. Prog. Phys. \textbf{80}, 076201 (2017).

\bibitem{Cheng:2021qpd}
H.~Y.~Cheng,
Charmed baryon physics circa 2021,
Chin. J. Phys. \textbf{78}, 324-362 (2022).


\bibitem{Ebert:2004ck}
D.~Ebert, R.~N.~Faustov, V.~O.~Galkin and A.~P.~Martynenko,
Semileptonic decays of doubly heavy baryons in the relativistic quark model,
Phys.\ Rev.\ D {\bf 70}, 014018 (2004).

\bibitem{Roberts:2008wq}
W.~Roberts and M.~Pervin,
Hyperfine Mixing and the Semileptonic Decays of Double-Heavy Baryons in a Quark Model,
Int.\ J.\ Mod.\ Phys.\ A {\bf 24}, 2401 (2009).

\bibitem{Branz:2010pq}
T.~Branz, A.~Faessler, T.~Gutsche, M.~A.~Ivanov, J.~G.~Korner, V.~E.~Lyubovitskij and B.~Oexl,
Radiative decays of double heavy baryons in a relativistic constituent three-quark model including hyperfine mixing,
Phys.\ Rev.\ D {\bf 81}, 114036 (2010).

\bibitem{Albertus:2010hi}
C.~Albertus, E.~Hernandez and J.~Nieves,
Hyperfine mixing in electromagnetic decay of doubly heavy $bc$ baryons,
Phys.\ Lett.\ B {\bf 690}, 265 (2010).

\bibitem{Bahtiyar:2018vub}
H.~Bahtiyar, K.~U.~Can, G.~Erkol, M.~Oka and T.~T.~Takahashi,
Radiative transitions of doubly charmed baryons in lattice QCD,
Phys. Rev. D \textbf{98}, 114505 (2018).

\bibitem{Qin:2021zqx}
Q.~Qin, Y.~J.~Shi, W.~Wang, G.~H.~Yang, F.~S.~Yu and R.~Zhu,
Inclusive approach to hunt for the beauty-charmed baryons~$\Xi_{bc}$, 
Phys. Rev. D \textbf{105}, L031902 (2022).


\bibitem{Eakins:2012fq}
B.~Eakins and W.~Roberts,
Heavy Diquark Symmetry Constraints for Strong Decays,
Int. J. Mod. Phys. A \textbf{27}, 1250153 (2012).

\bibitem{Xiao:2017udy}
L.~Y.~Xiao, K.~L.~Wang, Q.~F.~L\"u, X.~H.~Zhong and S.~L.~Zhu,
Strong and radiative decays of the doubly charmed baryons,
Phys. Rev. D \textbf{96}, 094005 (2017).

\bibitem{Xiao:2017dly}
L.~Y.~Xiao, Q.~F.~L\"u and S.~L.~Zhu,
Strong decays of the $1P$ and $2D$ doubly charmed states,
Phys. Rev. D \textbf{97}, 074005 (2018).


\bibitem{He:2021iwx}
H.~Z.~He, W.~Liang and Q.~F.~L\"u,
Strong decays of the low-lying doubly bottom baryons,
Phys. Rev. D \textbf{105}, 014010 (2022).

\bibitem{Mehen:2017nrh}
T.~Mehen,
Implications of Heavy Quark-Diquark Symmetry for Excited Doubly Heavy Baryons and Tetraquarks,
Phys. Rev. D \textbf{96}, 094028 (2017).

\bibitem{Ma:2017nik}
Y.~L.~Ma and M.~Harada,
Chiral partner structure of doubly heavy baryons with heavy quark spin-flavor symmetry,
J. Phys. G \textbf{45}, 075006 (2018).

\bibitem{Yan:2018zdt}
M.~J.~Yan, X.~H.~Liu, S.~Gonz\`alez-Sol\'\i s, F.~K.~Guo, C.~Hanhart, U.~G.~Mei\ss ner and B.~S.~Zou,
New spectrum of negative-parity doubly charmed baryons: Possibility of two quasistable states,
Phys. Rev. D \textbf{98}, 091502 (2018).

\bibitem{Chen:2022fye}
B.~Chen, S.~Q.~Luo, K.~W.~Wei and X.~Liu,
$b-$hadron spectroscopy study based on the similarity of double bottom baryon and bottom meson,
Phys. Rev. D \textbf{105}, 074014 (2022).

\bibitem{Song:2023cyk}
Q.~F.~Song, Q.~F.~L\"u and A.~Hosaka,
Bottom-charmed baryons in a nonrelativistic quark model,
Eur. Phys. J. C \textbf{84}, 89 (2024).


\bibitem{Nagahiro:2016nsx}
H.~Nagahiro, S.~Yasui, A.~Hosaka, M.~Oka and H.~Noumi,
Structure of charmed baryons studied by pionic decays,
Phys. Rev. D \textbf{95}, 014023 (2017).

\bibitem{Arifi:2021orx}
A.~J.~Arifi, D.~Suenaga and A.~Hosaka,
Relativistic corrections to decays of heavy baryons in the quark model,
Phys. Rev. D \textbf{103}, 094003 (2021).

\bibitem{Arifi:2022ntc}
A.~J.~Arifi, D.~Suenaga, A.~Hosaka and Y.~Oh, Strong decays of multistrangeness baryon resonances in the quark model,
Phys. Rev. D \textbf{105}, 094006 (2022).

\bibitem{Wang:2018fjm}
K.~L.~Wang, Q.~F.~L\"u and X.~H.~Zhong,
Interpretation of the newly observed $\Sigma_b(6097)^{\pm}$ and $\Xi_b(6227)^-$ states as the $P$-wave bottom baryons,
Phys. Rev. D \textbf{99}, 014011 (2019).

\bibitem{Liu:2019wdr}
M.~S.~Liu, K.~L.~Wang, Q.~F.~L\"u and X.~H.~Zhong,
$\Omega$ baryon spectrum and their decays in a constituent quark model,
Phys. Rev. D \textbf{101}, 016002 (2020).

\bibitem{Lu:2022puv}
Q.~F.~L\"u, H.~Nagahiro and A.~Hosaka,
Understanding the nature of $\Omega(2012)$ in a coupled-channel approach,
Phys. Rev. D \textbf{107}, 014025 (2023).


\bibitem{Hwang:2004cd}
D.~S.~Hwang and D.~W.~Kim,
Mass of $D^*_{sJ}(2317)$ and coupled channel effect,
Phys. Lett. B \textbf{601}, 137-143 (2004).

\bibitem{MartinezTorres:2011pr}
A.~Martinez Torres, L.~R.~Dai, C.~Koren, D.~Jido and E.~Oset,
The $KD$, $\eta D_s$ interaction in finite volume and the nature of the $D_{s^* 0}(2317)$ resonance,
Phys. Rev. D \textbf{85}, 014027 (2012).

\bibitem{Mohler:2013rwa}
D.~Mohler, C.~B.~Lang, L.~Leskovec, S.~Prelovsek and R.~M.~Woloshyn,
$D_{s0}^*(2317)$ Meson and $D$-Meson-Kaon Scattering from Lattice QCD,
Phys. Rev. Lett. \textbf{111}, 222001 (2013).

\bibitem{Lang:2014yfa}
C.~B.~Lang, L.~Leskovec, D.~Mohler, S.~Prelovsek and R.~M.~Woloshyn,
$D_s$ mesons with $DK$ and $D^*K$ scattering near threshold,
Phys. Rev. D \textbf{90}, 034510 (2014).

\bibitem{Ortega:2016mms}
P.~G.~Ortega, J.~Segovia, D.~R.~Entem and F.~Fernandez,
Molecular components in $P-$wave charmed-strange mesons,
Phys. Rev. D \textbf{94}, 074037 (2016).

\bibitem{Bali:2017pdv}
G.~S.~Bali, S.~Collins, A.~Cox and A.~Sch\"afer,
Masses and decay constants of the $D_{s0}^*(2317)$ and $D_{s1}(2460)$ from $N_f=2$ lattice QCD close to the physical point,
Phys. Rev. D \textbf{96}, 074501 (2017).

\bibitem{Albaladejo:2018mhb}
M.~Albaladejo, P.~Fernandez-Soler, J.~Nieves and P.~G.~Ortega,
Contribution of constituent quark model $c\bar{s}$ states to the dynamics of the $D_{s0}^*(2317)$ and $D_{s1}(2460)$ resonances,
Eur. Phys. J. C \textbf{78}, 722 (2018).


\bibitem{Wang:2021rjk}
J.~B.~Wang, G.~Li, C.~R.~Deng, C.~S.~An and J.~J.~Xie,
$\Omega_{cc}$ resonances with negative parity in the chiral constituent quark model,
Phys. Rev. D \textbf{104}, 094008 (2021).


\bibitem{Wang:2022aga}
W.~F.~Wang, A.~Feijoo, J.~Song and E.~Oset,
Molecular $\Omega_{cc}$, $\Omega_{bb}$, and $\Omega_{bc}$ states,
Phys. Rev. D \textbf{106}, 116004 (2022).

\bibitem{Wang:2023mdj}
Z.~Y.~Wang, C.~W.~Xiao, Z.~F.~Sun and X.~Liu,
Molecular-type $QQss \bar s$ pentaquarks predicted by an extended hidden gauge symmetry approach,
Phys. Rev. D \textbf{109}, 034038 (2024).


\bibitem{Lu:2016bbk}
Q.~F.~L\"u, T.~T.~Pan, Y.~Y.~Wang, E.~Wang and D.~M.~Li,
Excited bottom and bottom-strange mesons in the quark model,
Phys. Rev. D \textbf{94}, 074012 (2016).



\bibitem{Shi:2020qde}
Y.~J.~Shi, W.~Wang, Z.~X.~Zhao and U.~G.~Mei\ss{}ner,
Towards a heavy diquark effective theory for weak decays of doubly heavy baryons,
Eur. Phys. J. C \textbf{80}, 398 (2020).



\end{thebibliography}
\end{document}